\documentclass[12pt,superscriptaddress,eqsecnum,nofootinbib,color]{revtex4}
\usepackage{amsmath}
 \usepackage{color}

\newcommand{\tw}{\tilde{\omega}}
\newcommand{\bz}{\bar{z}}

\newcommand{\bea}{\begin{eqnarray}}
\newcommand{\eea}{\end{eqnarray}}
\newcommand{\beq}{\begin{equation}}

\newcommand{\eeq}{\end{equation}}

\begin{document}

\title{ GUP corrections to Dirac oscillator in the external magnetic field}

\author{Vishakha Tyagi \footnote{e-mail address:
\ \ tyagi.vishakha17@gmail.com.com }}
\affiliation{ Department of Physics,\\
Banasthali Vidyapeeth,\\
Rajasthan-304022, INDIA. \\
}
\author{Sumit Kumar Rai\footnote{e-mail address: sumitssc@gmail.com} }

\affiliation{ Sardar Vallabhbhai Patel College,\\
(Veer Kunwar Singh University, Ara),\\
Bhabua-821101, INDIA. \\
}
\author{Bhabani Prasad Mandal \footnote{e-mail address:
\ \ bhabani@bhu.ac.in, \ \ bhabani.mandal@gmail.com  } }
\affiliation{ Department of Physics,\\
Banaras Hindu University,\\
Varanasi-221005, INDIA. \\
}


\begin{abstract}
We have studied (2+1) dimensional  Dirac oscillator (DO) in  an external magnetic field in the framework of generalized uncertainty principle (GUP). We have calculated the perturbative corrections for first few energy levels. We show that the infinite degeneracy of lowest Landau level is partially lifted due to GUP correction and obtained a critical value of the magnetic field for which there is no GUP correction and the DO stops oscillating.
\end{abstract}
\maketitle
\bigskip

\section{Introduction}
 According to Heisenberg Uncertainty Principle (HUP) 
the position  and momentum cannot be measured simultaneously and arbitrarily accurate. In HUP, it is assumed that the position continuously varies from $-\infty$ to $+\infty$. However, with the emergence of various theories on quantum gravity and black hole physics, it has been predicted the existence of a certain minimum measurable length  called the Planck length ($l_{pl}\approx 10^{-35} m$ ) \cite{grme1,grme2,tadi,amci,koni,alva1,alva2}. The restriction $\Delta x \geq l_{pl} $, leads to the modification in HUP,  known as the generalized uncertainty principle (GUP) \cite{amci,magg1,gara,came,scar,magg2}. The modified commutation relations for the GUP which are consistent with string theories, black hole physics, doubly special relativity (DSR) etc. theories have been constructed in various ways \cite{alva1,scar,kema,mure,bhat,hike}. The simplest and commonly used commutation relation in GUP is 
\begin{eqnarray}
&&\left[x_i,p_j\right]=i\hbar\left(\delta_{ij}-a\left(p\delta_{ij}+\frac{p_ip_j}{p}\right)+a^2\left(p^2\delta_{ij}+3p_ip_j\right)......\right) \label{mgup}\\
&&\left[x_i,x_j\right]=0; \quad\quad \left[p_i,p_j\right]=0;
\end{eqnarray}
where $a=\frac{a_0}{M_{pl}c}=\frac{a_0 l_{pl}}{\hbar}$. $ M_{pl} $ is the Planck mass and $M_{pl}c^2\approx 10^{19}$ GeV is the Planck energy. The order of  $a_0$ is assumed to be unity.
 The generalized commutation relation in the equation (\ref{mgup}) can be expressed equivalently as \cite{alva2}
\begin{equation}
\left[x_i,p_j\right]=i\hbar\left(\delta_{ij}+a\delta_{ij}p^2+2ap_ip_j\right),\label{mgup1}
\end{equation}
where \ \ $p^2=\sum_{j=1}^3 p_jp_j$. The position and momentum operators in GUP further have been expressed as follows
\begin{equation}
x_i=x_{0i};   \quad\quad\quad p_i=p_{oi}\left(1-ap_0+2a^2p_0^2\right);\label{usm}
\end{equation}
where $p_{0i}$ is the momentum operator at low energies. $x_{0i}$ and $p_{0i}$ satisfies the usual canonical commutation relations $\left[x_{0i},p_{0i}\right]=i\hbar\delta_{ij}$. The position operator $x_i$ and momentum operator $p_i$ in GUP are  known as high energy, position and momentum operators respectively. $a$ dependent terms in Eq. (\ref{mgup1}) becomes crucial only at momentum (energy) scale comparable to Planck energy scales.

 GUP and its consequences have been extensively studied over last few years \cite{tadi, bhat,tadi1,scar2,maju,mazi,faiz,menc,pedr,fama,bafr,ghro,dehu,dey,defr,
chen,noka,dava,boss,boss2,dass3,oafr,peam,fakr,mafa,khno1,brcl,faiz3}. Klein-Gordon \cite{chen} and Dirac equations \cite{menc,noka,oafr,peam,fakr} have been modified under GUP corrections. Modified Dirac equation has found its application in graphene \cite{iopa}. GUP has found wide applications in various non-relativistic quantum systems such as a particle in a box, simple harmonic oscillator, Landau levels \cite{alva2,dava,mafa,caob}, coherent and squeezed states \cite{ghro,dehu,dey,defr}, PT symmetric non-Hermitian systems \cite{pedr,fama,bhat,bafr}, Schwinger's model of angular momentum \cite{saku,vemi} etc.  GUP has led to the discretization of time which resembles a crystal lattice in time known as time crystals \cite{fakh}.
In the recent years, quantum gravity effects have been studied in the field of astrophysics and cosmology such as black holes and its thermodynamics \cite{gesu1,ong,chun,khno,fakh1,gadu,gesu}, mass-radius relation of white dwarfs \cite{mana},  gravitational field  \cite{deda,zhfa1}, deformation of Wheeler-DeWitt equation \cite{faiz2,gafa}, de-Sitter universe \cite{moal,alfa}, non-singular and cyclic universe \cite{saha} etc. Its effects have also been studied  in the gravitational wave event \cite{feya}. The consequences of GUP has been extended to field theories as well \cite{gium,vaal,fama1,zhfa,fats}such as non-local field theories \cite{gium}, no cloning theorem \cite{vaal} and Lifshitz field theories \cite{fama1}, supersymmetric field theories \cite{zhfa} and topological defects in deformed gauge theory \cite{fats}. The path integral approach to quantization has also been studied \cite{prfa}. The supersymmetry breaking has been found as a new source of the GUP and the relation between GUP and Lee-Wick field theories has been established \cite{faiz1} . 

Ito {\it et al} \cite{itmo} studied Dirac equation by adding a  linear harmonic potential $-imc\omega\beta{{\mbox{\boldmath $\alpha $}}}{\bf \cdot r}$  and later it was found that in the non-relativistic limit, it reduces to a harmonic oscillator with a strong spin orbit coupling and so referred as Dirac oscillator by  Moshinsky and Szczepaniak \cite{mosz}.  The Dirac Hamiltonian for a free particle with a linear harmonic potential term is written as
\begin{equation}
H=c {\mbox{\boldmath $\alpha $}}\cdot\left({\bf p}-\iota m\omega\beta{\bf r}\right)+\beta m c^2.
\end{equation}
 ${\bf \alpha}$ and $\beta$ are the usual Dirac matrices. $m,c$ and $\omega$ are the rest mass of the particle, speed of light and oscillator frequency respectively. Dirac oscillator has wide applications in various branches of physics \cite{lima,lima1,lima2,lima3,lima4,lima5,lima6,lima7,lima8,lima9,lima10,lima11,lima12,lima13,lima14,lima15,noc,noc1}.
In the recent years, Dirac Oscillator has been studied in the framework of GUP correction. It has been shown that Dirac oscillator in magnetic field with GUP corrections in ordinary quantum mechanics has a single left-right chiral quantum phase transition. The existence of GUP correction modifies the degeneracy of states and some of the states do not exist in the ordinary quantum mechanics limit\cite{menc,mepa}.

In this present letter, we solve the (2+1-D) Dirac equation in the presence of external magnetic field with GUP correction using perturbation theory and investigate the behaviour of the system at critically high magnetic field. For that purpose, we first calculate the modified Hamiltonian for DO in the framework of GUP upto first order corrections.  The additional momentum dependent term arises due to GUP correction, then dealt with perturbation. Degeneracy of Landau levels are lifted due to GUP corrections.We  obtain a critical value of magnetic field for which no GUP corrections and the dynamics of DO stops. 

In an earlier work \cite{mave}, we showed that (2+1)-D  Dirac oscillator in an external magnetic field  can be  mapped onto the Dirac oscillator without magnetic field but with the reduced angular frequency given by the following equation \cite{mave,rama}
\begin{equation}
H= c{\mbox{\boldmath$\alpha$}}\cdot ({\bf p} -\iota m\tw\beta {\bf r}) + \beta m c^2, \quad {\mbox where}\;\; \tw =\omega - \frac{\omega_c}{2}.
\label{dowm}
\end{equation}
where the angular frequency $\tw$ is reduced by half the value of cyclotron frequency $\omega_c=\frac{|e|B}{mc}$. The momentum $\bf p$ in the above equation is the high energy momenta. The magnetic field $ B$ is taken along the $z$-direction for convenience.
${\bf A}$ is the vector potential chosen in the symmetric gauge as ${\bf A}= ( -\frac{B}{2}y,
\frac{B}{2}x, 0)$ and 
$e$ is the charge of the Dirac oscillator.
 The Dirac Hamiltonian in the Eq. (\ref{dowm}) can be expressed in the  matrix form as  
\begin{equation}
H=\left ( \begin{array}{c c} mc^2 & 2c p_z+\iota m \tw c \bar{z} \\2c p_{\bar{z}}+\iota m \tw c z & -mc^2 \end{array} \right )
\label{nham},
\end{equation}
where the momenta $p_z$ and $p_{\bar{z}}$  are defined as follows
\begin{eqnarray}
 p_z &=&-i\hbar\frac{d}{dz} = \frac{1}{2}(p_x-ip_y) \nonumber ,\\
 p_{\bz}&=&- i\hbar\frac{d}{d\bz} = \frac{1}{2}(p_x+ip_y),
\label{pz1} 
\end{eqnarray}
with $ \left[z,p_z\right]=i\hbar=\left[\bar{z},p_{\bar{z}}\right]$, $\left[z,p_{\bar{z}}\right]=0=\left[\bar{z},p_z\right]    $ and hence $p^2 =4p_zp_{\bar{z}}$.
The GUP modified Dirac oscillator in Eq. (\ref{dowm}) upto the first order in $a$ can be expressed as \cite{sava}
\begin{eqnarray}
H\psi &=& \left(c{\mbox{\boldmath$\alpha$}}\cdot \left({\bf p} -i m\tw\beta {\bf r}\right) + \beta m c^2\right)\psi , \nonumber \\
&=&\left[c\left({\mbox{\boldmath$\alpha$}}\cdot {\bf p_0}-a({\mbox{\boldmath$\alpha$}}\cdot {\bf p_0})({\mbox{\boldmath$\alpha$}}\cdot {\bf p_0}) -{\mbox{\boldmath$\alpha$}}\cdot (i m\tw \beta {\bf r})\right) + \beta m c^2\right]\psi , \\
&=&E\psi \nonumber 
\end{eqnarray}
We drop the subscript from the position and momentum operator and henceforth all the variables are low energy variables which satisfy usual commutation relations.

The usual Hamiltonian in equation ({\ref{dowm}) with the GUP correction  can be expressed as the following Hamiltonian
\begin{eqnarray}
H&=&\left ( \begin{array}{c c} mc^2-acp^2 & 2c p_z+im \tw c \bar{z} \\2c p_{\bar{z}}+i m \tw c z & -mc^2-acp^2 \end{array} \right ) \nonumber\\
&=&\left ( \begin{array}{c c} mc^2 & 2c p_z+i m \tw c \bar{z} \\2c p_{\bar{z}}+i m \tw c z & -mc^2 \end{array} \right )-a\left ( \begin{array}{c c} cp^2 & 0 \\0 & cp^2 \end{array} \right )
\label{gupham},
\end{eqnarray}
where the two component $\left|\psi \right\rangle= \left(\begin{array}{c c} \left|\psi_1\right\rangle 
\\ \left|\psi_2\right\rangle  \end{array} \right )$.
The above Hamiltonian can be expressed as 
\begin{equation}
H=H_0+H^\prime
\end{equation}
where $H_0$ is the usual Hamiltonian. Dirac equation for $H_0$ has been solved by writing  $\left|\psi \right\rangle= \left(\begin{array}{c c} \left|\psi_1\right\rangle 
\\ \left|\psi_2\right\rangle  \end{array} \right )$ and 
\begin{equation}
H^\prime = -\left ( \begin{array}{c c} acp^2 & 0 \\0 & acp^2 \end{array} \right )
\end{equation}
is dealt with perturbative techniques.
The normalized negative and positive energy states for $H_0$ are given by \cite{mave}
\begin{equation}
\left|\psi_n^\pm\right\rangle =c^\pm_n \left|n;\frac{1}{2}\right\rangle +d^\pm_n\left|n-1;\frac{-1}{2}\right\rangle;  \label{state}
\end{equation}
where the coefficients $ c_n^\pm =\pm\sqrt{\frac{E_n^+ \pm mc^2}{2E_n^+}} $ and $ d_n^\pm =\sqrt{\frac {E_n^+ \mp mc^2}{2E_n^+}} $. In the equation (\ref{state}), the notation adopted is $ \left.|n, \frac{1}{2} m_s\right \rangle \approx \psi_n \left( z, \bar{z}\right) \Phi_{m_s}$. $\psi_n \left( z, \bar{z}\right)$ is the spatial part of the wave function whereas $\Phi_{m_s}$ is the spin part of the wave function. $n$ is the eigenvalue of for the number operator $a^+ a$ and $m_s=\pm 1$ are the eigenvalues of the spin operator $\sigma_z$.
When the  creation and annihilation operators are defined as \cite{mave}
\begin{eqnarray}
a&=&\frac{1}{\sqrt{m\tw\hbar}}p_{\bar{z}}-\frac{i}{2}\sqrt{\frac{m\tw}{\hbar}}z \nonumber \\
a^\dagger &=& \frac{1}{\sqrt{m\tw\hbar}}p_{z}-\frac{i}{2}\sqrt{\frac{m\tw}{\hbar}}\bar{z} \label{cran}
\end{eqnarray} 
The relativistic Landau levels due to the unperturbed Hamiltonian $H_0$ is given by \cite{mave}
\begin{equation}
E^\pm_n=\pm mc^2 \sqrt{1+\frac{4\hbar\tw}{mc^2}n}, \quad\quad n=0,1,2,3 .........
\end{equation}
For $n=0$, the state is $|\psi_0^{\pm}\rangle = c^{\pm}_0|0,\frac{1}{2}\rangle$.
The GUP corrected part $H^\prime$ can be dealt with perturbative techniques as it is proportional to $a$ which is extremely small.
Using Eq. (\ref{pz1}) and Eq. (\ref{cran}), $p^2$ can be expressed as 
\begin{equation}
4p_zp_{\bar{z}}=2m\tw\hbar\left[a^\dagger a+a a^\dagger-\frac{m\tw}{2\hbar}z\bar{z}+\frac{\hat{L}_z}{\hbar}\right]
\end{equation}



First order correction to the  $n^{th}$  state is given by 
\begin{equation}  
\left\langle\psi_n | 4p_z p_{\bar{z}}|\psi_n\right\rangle = 2m\tw\hbar\left\langle\psi_n | a^\dagger a+a a^\dagger|\psi_n\right\rangle -{(m\tw)}^2 \left\langle\psi_n | z {\bar{z}}|\psi_n\right\rangle + 2m\tw\left\langle\psi_n | \hat{L}_z|\psi_n\right\rangle         \label{nstate}
\end{equation}
It is difficult to  obtain a general expression for the correction in the $n^{th}$ level due to degeneracy of the levels $n\geq 2$.

Ground state energy correction arising out of the perturbative Hamiltonian $H^\prime$ for $n=0$ in Eq.(\ref{nstate}) can be obtained 
\begin{equation}
\left\langle\psi_0| acp^2|\psi_0\right\rangle = ac\left[ 2m\tw\hbar\left\langle\psi_0 | a^\dagger a+a a^\dagger|\psi_0\right\rangle -{(m\tw)^2} \left\langle\psi_0 | z {\bar{z}}|\psi_0\right\rangle \right]
\end{equation} 
For the ground state $l=0$, hence no contribution from the last term. Space part of the ground state wave function  $\psi_0 = C_0\exp^{-\frac{m\tw}{2\hbar}z\bar{z}}$ is used to calculate the middle term in the Eq. (\ref{nstate}). The correction to the ground state is then given by 
\begin{equation}
E_0^\prime = -acm\tw\hbar
\end{equation}
Using equations (\ref{state}) and (\ref{nstate}), one can obtain the energy correction for the first excited state as
\begin{equation} 
\left\langle\psi_1| 4p_z p_{\bar{z}}|\psi_1\right\rangle = -\frac{5 m\tw\hbar}{2}- 2m\tw\left\langle\psi_1 | \hat{L}_z|\psi_1\right\rangle;
\end{equation}
\begin{equation}
E_1^\prime =-\frac{5 ac m\tw\hbar}{2}
\end{equation}

Second excited state is 4-fold degenerate, therefore we need to use degenerate perturbation theory to calculate the correction for these states. We write the matrix  as 
\begin{eqnarray}
&=&\left ( \begin{array}{c c c c} \langle\psi_{200}|H^\prime |\psi_{200}\rangle & \langle\psi_{200}|H^\prime |\psi_{210}\rangle &\langle\psi_{200}|H^\prime |\psi_{211}\rangle &\langle\psi_{200}|H^\prime |\psi_{21-1}\rangle  \\ \langle\psi_{210}|H^\prime |\psi_{200}\rangle  & \langle\psi_{210}|H^\prime |\psi_{210}\rangle  & \langle\psi_{210}|H^\prime |\psi_{211}\rangle &\langle \psi_{210}|H^\prime |\psi_{21-1}\rangle \\ \langle \psi_{211}|H^\prime |\psi_{200}\rangle & \langle\psi_{211}|H^\prime |\psi_{210}\rangle &\langle\psi_{211}|H^\prime |\psi_{211}\rangle &\langle\psi_{211}|H^\prime |\psi_{21-1}\rangle  \\ \langle\psi_{21-1}|H^\prime |\psi_{200}\rangle & \langle\psi_{21-1}|H^\prime |\psi_{210}\rangle &\langle\psi_{21-1}|H^\prime |\psi_{211}\rangle &\langle\psi_{21-1}|H^\prime |\psi_{21-1}\rangle    \end{array} \right ) \nonumber \\
&=&-\frac{acm\tw\hbar}{2} \left ( \begin{array}{c c c c} 11 & -5 &-5 &-5 \\ -5 & 11 &-5 &-5 \\ -5 & -5 &13 &-5 \\ -5 & -5 &-5 &9  \end{array} \right ) 
\end{eqnarray}
The perturbative energy corrections are obtained by diagonalizing the above matrix. The energy eigenvalues obtained are 
\begin{eqnarray}
E_2^{\prime (1)}&=&-8.7308 acm\hbar\tw; \qquad E_2^{\prime (2)}=-8acm\hbar\tw; \nonumber \\ 
 E_2^{\prime (3)}&=&-7.3192acm\hbar\tw; \qquad E_2^{\prime (4)}=2.05acm\hbar\tw ;
\end{eqnarray}
The corresponding eigenvectors are as follows
\begin{equation}
\left(\begin{array}{c} {|\psi_2\rangle}_1 \\ {|\psi_2\rangle}_2 \\{|\psi_2\rangle}_3 \\{|\psi_2\rangle}_4  \end{array} \right)=        
\left ( \begin{array}{c c c c} 2.36839 & 2.36839 &-6.42909 &1 \\ -1 & 1 &0 &0 \\ -0.468884& -0.468884 &-0.189917&1 \\ 0.900498 & 0.900498 &0.819005 &1  \end{array} \right ) \left(\begin{array}{c} {|\psi_{200}\rangle} \\ {|\psi_{210}\rangle} \\{|\psi_{211}\rangle} \\{|\psi_{21-1}\rangle}  \end{array} \right)
\end{equation}

In conclusion, we have studied the consequences of generalized uncertainty principle in modified (2+1) dimensional Dirac oscillator in the presence of magnetic field. The perturbative corrections upto first few energy levels have been explicitly calculated. The degeneracy of Landau levels are partially lifted due to GUP  corrections. An interesting situation occurs for a critical value of magnetic field i.e $B=\frac{2\omega mc}{|e|} $ when $\tw=0$. The Dirac oscillator stops oscillating even with the GUP corrections.

Recently the discreteness of space is shown to arise in Schwarzschild metric due to the GUP correction but still we do not have it in most generic curved space time \cite{abut,barr}. Dirac oscillator in the context of gravity rainbow has been studied where the modified energy levels have been obtained \cite{bamo}. It would be very  interesting to extend our approach to the most generic curved-space time and to the gravity rainbow for future studies.
 
{ \bf Acknowledgments}: 
 BPM acknowledges the support from MATRIX project (Grant No. MTR/2018/000611), SERB, DST
Govt. of India.

\end{document}